# Why developers cannot embed privacy into software systems?

An empirical investigation

Awanthika Senarath
University of New South Wales
Canberra, ACT
a.senarath@student.unsw.edu.au

Nalin A. G. Arachchilage
University of New South Wales
Canberra, ACT
nalin.asanka@adfa.edu.au

## ABSTRACT

Pervasive use of software applications continue to challenge user privacy when users interact with software systems. Even though privacy practices such as Privacy by Design (PbD), have clear instructions for software developers to embed privacy into software designs, those practices are yet to become a common practice among software developers. The difficulty of developing privacy preserving software systems highlights the importance of investigating software developers and the problems they face when they are asked to embed privacy into application designs. Software developers are the community who can put practices such as PbD into action. Therefore identifying the problems they face when embedding privacy into software applications and providing solutions to those problems are important to enable the development of privacy preserving software systems. This study investigates 36 software developers in a software design task with instructions to embed privacy in order to identify the problems they face. We derive recommendation guidelines to address the problems to enable the development of privacy preserving software systems.

## KEYWORDS

Software Development, Usable Privacy, Privacy Practices

## 1 INTRODUCTION

With the excessive use of connected services, such as mobile applications, Internet of things and online networks, personal data of users are accessed, stored, processed and shared in ways users cannot control [27]. Users find it increasingly difficult to understand how their data is shared and processed once they disclose data into an online network, such as Facebook. However, this is not the user's fault. Software systems that access and process user data should be designed with privacy, so that users' privacy is not compromised when they interact with these systems [5]. For this, software developers are expected to embed privacy into the software systems they design [25, 26].

In order to instruct software developers to embed privacy into software system designs, there are various privacy practices that are well established and widely known, such as Fair Information Practices (FIP) [29], Privacy by Design (PbD) [5] and Data Minimization (DM) [26]. The principle of DM states that data should only be collected if they are related to the purpose of the application, and should be processed only for the purpose for which they were collected. FIP says that users should have access and control over their data even after they disclose data into a system [29]. However, do developers really follow these practices when they design software systems? If software systems are designed to only process data for the purpose for which they were collected, and if users are given control over their data, we would not see the recent privacy scandal of Facebook, where Cambridge Analytica accessed and processed personal data of 50 million users without the users' consent. Facebook's response to the scandal saying that they would take strict measures to limit developers' access to user data and ban developers who disagree to privacy audits indicate that developers privacy engagement may have a significant effect on such issues [10].

Software systems continuously failing user privacy suggests that the domain of software developers who can put practices such as PbD into action should be investigated to understand why they fail to design software systems with privacy [3]. Developers are either not considering these practices at all when they are asked to embed privacy into a software design or the way they implement these privacy practices in software systems is not satisfactory. Because of this, software applications end up being developed with privacy vulnerabilities, and this not only results in users losing their privacy, but also in software development organizations losing their reputation and market value. For example, on the following day of the Cambridge Analytica incident, Facebook's shares lost its market value by $58 billion [10]. Therefore, it is important that the issues developers face when they attempt to embed privacy into software applications are understood [3, 14]. Addressing these problems would enable the development of privacy practices that can effectively guide software developers to embed privacy into the software applications they design.

For this, in this research, we investigate *What are the issues developers face when they attempt to embed privacy into software systems?*. We assigned a software design task to 36 software developers and asked them to embed privacy into their design. Through a post task questionnaire we then investigated how participants embedded privacy into the software designs and what issues they faced when they attempted embedding privacy into the software design. Our findings revealed that most developers lack formal knowledge on privacy practices. such as PbD, and that they try to integrate privacy features into software designs without much understanding. Participants also found privacy to contradict with system requirements and they had trouble evaluating whether they have successfully embedded privacy into a design. Based on these findings, we provide recommendation to help the development of practical and effective privacy practices to guide software developers to embed privacy into the software systems they design.

## 2 RELATED WORK

The effects of developers privacy engagement towards the resulting privacy in software systems has become an interest of privacy researchers recently. There has been several recent studies that attempt to investigate how developers engage with privacy requirements when they develop software systems.

Most of these studies focus on organizational privacy practices. For example, Hadar et al. [14] revealed that organizational culture and policies play a significant role both positively and negatively in encouraging developers to consider privacy in their work. Similarly, both Sheth et al. [28] and Jain et al. [15] emphasize that organizations should setup policies and guidelines to guide software developers to embed privacy into the software systems they design. However, so far no one has investigated how organizations should setup these privacy policies, and what issues they need to address in these policies, so that they can effectively guide software developers to embed privacy into the software systems they design.

Particularly focusing on the attitudes of software developers towards privacy, Ayalon et al. [3] say that developers reject privacy guidelines that do not follow existing software frameworks. Similarly, Sheth et al. [28] say that developers find it difficult to understand privacy requirements by themselves when they develop software systems. Oetzel et al. [21] have said that developers require significant effort to estimate privacy risks from a user perspective. These investigations [3, 28] suggest that developers face problems when they attempt to embed privacy into software applications. It is said that through effective guidelines and practices developers could be nudged to make privacy preserving choices when they develop software applications [15]. However, for this, privacy guidelines and practices need to address the actual problems developers face when they try to embed privacy into software systems [15, 22, 31]. Nevertheless, so far no study has been conducted to identify the exact issues developers face when they attempt to embed privacy into software applications.

In this research, we seek to empirically investigate the issues software developers face when they attempt to embed privacy into software systems. We focus on developers with industry experience in end user software application development. By this, we observe those who are asked to embed privacy into the applications they design. Through the grounded theory approach [18] we identify the issues developers face when they attempt to embed privacy into a software application design and then we relate these issues into guidelines that should be considered when establishing privacy practices for software developers.

## 3 METHODOLOGY

We designed this study to investigate the issues software developers face when they attempt to embed privacy into software systems. The complete study design was approved by the ethic committee of the University of New South Wales, Australia.

Following similar security and privacy related studies that investigate software developers [1, 33], we decided to conduct our study remotely, in order to encourage uncontrolled behavior in developers. Previous work has shown that if special attention is given to designing the task, remote studies could be used to investigate the behaviors of participants in tasks ranging from simple surveys to prototype development [1, 19]. In this study we assigned a software application design task to the participants with a post-task questionnaire. This approach was used to minimize the limitations in survey methodologies that arise due to participants' memorability and recalling capacity of their usual actions when they answer questions [2]. By asking questions after the participants completed a design activity, we do not ask them about their general behavior. Rather we ask them about the task they completed. Therefore, rather than expressing what they think they would do, or what they think they did in the past, participants express what they did in the task. This way we aim to identify the exact issues developers face when they attempt to embed privacy into application design. Both the questionnaire and the design task were carefully evaluated and fine tuned through a pilot study with two experienced developers known to the authors and not related to the study.

We used a health application as the scenario for this study as health data is highly sensitive and hence regulated in many regions around the world [20]. Therefore, privacy is a critical concern in health related software applications and thus this provides an ideal setup for our experiment [12, 17]. Following is the scenario we provided to the participants.

*You are required to design a web-based health-care application that allows remote consultation with medical professionals, general practitioners and specialists, for a payment. Users should be able to browse through a registered list of medical professionals and chat (text/video) with them on their health problems for advice. Doctors and health-care professionals can register on the application to earn by providing their expertise to users. The application is to be freely available on-line (desktop/mobile) and charge for individual consultations. You may consider advertising and data sharing with third parties such as insurance providers and hospitals to earn from user data. Consider user privacy as a requirement in this design.*

We recruited professional developers with industry experience from Github in order to obtain a diverse sample of participants, which is difficult to achieve when recruiting participants through local software development firms or university [33]. This is a widely accepted and commonly used method for participant recruitment for privacy (and security) research that require software developers [1]. We selected active Github committers from Java and PHP repositories as we were looking for end user application developers. We sent 6000 invitation emails with information on what developers had to do in the study, the duration of the study and the type of information we collect in the study. We also stated that the participation was voluntary and that they could withdraw from the study at any time. All participants were given an amazon gift voucher of USD $15 as a token of appreciation. We received 118 responses to the invitations and to these developers we sent a second email with instructions to participate. This email had the ethic consent form, participant information sheet and instruction guidelines sheet attached. We first asked the participants to read and understand the participant information sheet and then sign and send us the ethic consent form which gave us the right to collect and store the study results. From the 118 developers who expressed their interest, only 37 participants agreed to participate in the study, from which one entry was removed due to lack of quality.

First, participants were asked to read the scenario for the application and design the application. Participants were explicitly asked to embed privacy into the design as we were focusing on conscious behavior. In designing the application, participants were first asked to identify the privacy impact for stakeholders of the application (ex: doctors, patients etc.). With this we aimed to observe how participants conduct a Privacy Impact Assessment (PIA) [21]. Then, we asked them to list user data they wanted to collect for the application and give reasons for selecting each data element to observe how they apply DM when selecting data. Next, participants were asked to draw the information flow diagram for the application with privacy. With this, we aimed to observe the use of PbD and FIP practices to embed privacy into a software design, because it is said that information flow diagrams can be used to decide when and where to use privacy practices such as FIP and PbD [16, 35]. Finally, we asked for a database diagram to observe how participants would use database level privacy techniques as separation, aggregation and encryption [24]. By each of these steps we aimed to investigate the understanding of developers on the concepts behind well established privacy practices such as DM and PbD, and the way they incorporate these practices into their work when they are asked to embed privacy into a design. Through this we expected to identify the practical problems developers encounter when they attempt to embed privacy into a software design. Participants were given the freedom to use softwares or pen and paper in their designs. We provided examples for each of the diagrams we requested (data-base sketches, information flow diagrams) in the instruction guide.

Once the design activity was completed, participants were asked to share the diagrams with us through email or any other preferred method and fill out the post-task questionnaire. The questionnaire was designed to investigate the problems participants faced when they attempted embedding privacy into the designs. For example, we first asked the participants "Did you use the concept of DM when you decided to collect data?". If they said yes, we asked them to explain how they used it in the design. If they said no, we asked the participants "Why didn't you use DM in your design?". Finally, we also asked them if they knew the concept of DM and whether they had used it before in order to test their formal knowledge on the concept. The direct questions with yes/no answers in the questionnaire were designed following the methods for theory testing and modification requirements in quantitative research [23]. Open ended questions that encouraged descriptive answers were designed with the aim to build theories on the issues developers face when they embed privacy into a software application [23].

When analyzing the results, we first checked participants' answer to the yes/no questions to test whether they had consciously followed privacy practices, such as DM, when they were embedding privacy into their designs. Then we analyzed the designs to observe how participants had embedded privacy into the designs. This also helped us to investigate if the participants had used the concepts behind the privacy practices such as DM and relate it to what they claimed in the questionnaire. Then we analyzed participants' formal knowledge on the privacy practices to see how developers' knowledge affected the way they embed privacy into the applications they design. Finally, from their answers to the open ended questions we identified the issues they faced when they attempted to embed privacy into the application designs. The descriptive answers were coded in Nvivo [11], by using a self coding scheme [34]. First all answers were summarized and two coding schemes were generated by two researchers. As these two were similar and interchangeable, one coder then coded all answers using one coding scheme. These codes were then categorized and analyzed in several iterations following the grounded theory approach [7] to identify the issues (thematic analysis). We made use of the observations we made from the design diagrams and yes/no questions to guide us in extracting the issues through the codes. Due to space constraints, here we only present the final set of issues identified.

## 4 RESULTS AND DISCUSSION

We identified a total of 5 issues after merging 18 codes through 3 iterations. We reached theoretical saturation in the qualitative analysis at 17 participants. Table 01 shows the five themes that represent the issues we identified from the study. Next, we discuss these issues in detail with our observations.

*4.1.1 Participants complained that privacy contradicts system requirements.* Most participants identified privacy as a non-functional requirement (83.3%). Whether privacy is a functional requirement or not would be a subject for another debate, however, because participants identified it to be non-functional they considered system requirements should be given priority over privacy requirements. It is true that privacy requirements mostly contradict system requirements when developing software applications [9]. However, as suggested by the PbD principles, achieving privacy in a software design should not necessarily mean that the application's functionality is compromised [5, 6]. Nevertheless, these are complex decisions that require management support, and in order to get management support developers need to raise these concerns to the management. If developers themselves make the decision that system requirements should get precedence, privacy concerns are likely to not get implemented at all. Our findings indicate that when instructed to consider privacy in the design, participants do not give priority to privacy requirements. Because of this, some participants had purposely disregarded some privacy techniques altogether. For example, P3 said, *developer has to give priority to the client/business requirements, if I want to identify polish immigrants with diseases, I need last names or ethnic backgrounds. Would the anonymization process retain enough of a correlation there?*. This participant had disregarded both anonymization and pseudonymization in the design, which are important privacy preserving techniques in data collection [28]. Similarly some participants claimed that they disregard privacy requirements as a designer because privacy and system requirements mostly don't get along. Participants did not know how to achieve privacy together with system requirements because they had trouble relating privacy techniques into privacy requirements, which leads to the next problem we identified.

Table 1: Issues participants faced when embedding privacy into the designs

| Issue | Representative Quotes | Coverage (out of 36) |
|---|---|---|
| Contradiction | Requirements in this design contradict with privacy requirements | 27.6% (10) |
| Relating requirements to practice | How can I implement the fair information practices?, embedding privacy is too complex | 33.3%(12) |
| Assurance | I tried to use the theories in the design but I'm not sure if I really did, I think I did, I don't know if I have done it right | 38.8%(14) |
| Personal opinion | storing raw data in the db does not affect the privacy | 16.7%(6) |
| Lack of knowledge | I did not use the this privacy theory because I don't know them, I haven't heard of PIA | 19.4% (7) |

*4.1.2 Participants had trouble relating privacy requirements into privacy techniques.* From the information flow diagrams and database designs of the participants we could observe that some had attempted to incorporate privacy techniques such as anonymization and data separation when processing and storing data in the application design (Aggregation = 23, Separation = 25, Anonymization = 20, Pseudonimization = 11, Data Expiry (Storage period) = 18, Encryption = 22, out of 36). However, these numbers are not satisfactory. For example, 16 out of 36 developers failing to consider anonymization in the design, when they were asked to embed privacy into the design suggests that developers find it difficult to map technical features into privacy requirements. Oetzel et al. [21] have also stated that developers do not know privacy requirements such as DM could be fulfilled by technical implementations such as anonymization and pseudonimization through which user's identity is separated from their personal data. This lead to participants not incorporating privacy techniques adequately into their designs.

Because of the difficulty to relate privacy requirements into techniques they could implement, participants also claimed that application of privacy into a design was complex. For example, P5 said, *The principles are complex and lengthy, how can I implement them?*. Previous work has also challenged privacy practices such as PbD for being complex and too theoretical to be used within the software development processes [13, 21, 30, 32]. These privacy practices are mostly generated through social and legal research [4, 21], in order to broadly capture a vast amount of privacy requirements and present them in an abstract form, such as the *respect for user privacy* principle in PbD. The implementation of these principles requires a lot of work from a developer's end. However, participants demonstrated frustration when they had to make decisions, such as when to encrypt, when to anonymize and to which level privacy should be considered while embedding privacy into the application. Software developers are from a technical background and therefore it was difficult for them to grasp these *soft decisions*. For example, P7 said *I think that striving for perfection has no borders, I should know to which extent...*, because these are not measurable decisions that can be quantified, such as coding errors, which developers are used to [21]. P32 said, *A developer just needs to get stuff out the door so they can eat. Although I take privacy and security very importantly, in my experience it is very difficult to make it work like this*.

*4.1.3 Participants had trouble verifying their work.* One participant (P14) who claimed s/he had practiced PbD in the answers to the post-task questionnaire had not encrypted the contact details of the participant. This participant had not anonymized users in the database either. From such a design users' contact details could be easily leaked through a database hack, which could lead to a privacy and a security breach of system users. However, since there was no criteria for evaluation in PbD, after partially applying PbD the participant claimed that s/he used PbD in the design. Spiekermann [30], has said that guidelines such as PbD, give a developer the feeling that all s/he has to do is to select some privacy techniques and embed them into an application, whereas in reality the expectations of the guidelines are deep and challenging. However, without a criteria for evaluation developers cannot say whether they have adequately addressed these expectations in a design.

Because of this, almost half of those who claimed they used privacy practices such as DM and PbD had difficulty in describing how they applied the practice in the design. For example, most participants claimed they used the concept of DM (75%), however, when they were asked to explain, they said that they were not sure if they did it right. This was further accentuated by P11 who said that *I used data minimization, but I don't think I have made a good one, I tried to make it work*, and also by P32, *All of the questions here asked if I understand specific aspects and used them; without [evaluating] the way I used them, one would have a hard time saying yes*. These remarks suggest that participants had difficulty ensuring that they have practiced a privacy theory right. Because of this some participants attempted to use concepts such as aggregation for data storage and had given up later on (P15 & P8) in the design as they had no feedback to know what they were doing was right. P11 said *One needs to formalise these processes so that we \*know\* it is being done right*. Need for evaluation and demonstration of assurance was also coined by the ENISA guidelines for Privacy and Data Protection by Design [8]. Evaluation and feedback would make it easy for a third party or for the practitioner himself to verify whether or not s/he has practiced a guideline successfully and also reduce developer's personal opinions affecting the way they embed privacy into software applications, which is the next issue we identified.

*4.1.4 Participants personal opinions affected the way they embedded privacy into the design.* Some participants in our study said that techniques like encryption and data expiry are not important for privacy. P7 said *I don't see how plain text storage of these data in secure database affect user privacy* and P23 said *expiring data [deleting data after a period] is not important*. P7 said *These are all general information that does not violate much of user privacy*. P7 had not incorporated any privacy technique at database level and P23 had not considered data expiry because they thought these techniques were not important. Developers' personal opinions that go against best practices for embedding privacy into software applications hence affected the way developers embedded privacy into the application design. Ayalon et al.[3] have also claimed that developers' personal privacy preferences significantly affect the way they embed privacy into software applications. This happens because developers are not privacy experts and they lack knowledge and formal education on privacy practices and data protection requirements.

*4.1.5 Participants lacked knowledge on privacy practices.* Surprisingly most participants had no formal education or knowledge on well known privacy theories that are widely discussed and appreciated within the community of privacy researchers. Figure 01 shows the knowledge and the trainings the participants had had on four well known privacy practices.

The number of participants who had not even heard of these theories surprised us. Almost 70% of the participants did not have experience using at least one of the privacy practices we investigated (DM - 30, FIP - 31, PbD - 31, PIA - 32, PET - 26 out of 36). All participants in our study were either end user application developers, or had past experience in end user application development (least experienced participant had1 year of experience). Therefore, majority of them having no experience with commonly used privacy practices suggests the lack of formal education in developers on privacy. When observing the designs we noticed that participants who had previous experience and who knew the privacy practices had better privacy in their designs compared to those who tried to use the concepts behind those practices without formal education. For example, 10 participants said they did a PIA when doing the stakeholder analysis. However, from their submissions we observed that only 2 out of those 10 had conducted a successful PIA and both of them had used PIA before and had been trained on PIA. Nevertheless, we also observed participants who had previous experience with PIAs, that had incomplete PIAs in their designs where they had either not identified all stakeholders, or disregarded privacy impact on some stakeholders. For example, one participant said that *Doctors should not have any privacy concerns here, as they are the service providers.* Therefore, when educating developers on privacy practices, they should be trained and educated properly. Otherwise, incorrect application of practices could give developers a false sense of satisfaction that their software products are privacy inbuilt, whereas in reality the applications have incomplete privacy features built in.

## 5 RECOMMENDATIONS

We recommend the following guidelines to address the issues we identified through the experiment.

- ï Privacy guidelines should be simple, straightforward and explicit as developers have trouble executing soft decisions.
- ï Developers should be given formal education on privacy practices as developers' lack of knowledge affects their personal opinion which interferes the way they embed privacy into designs.
- ï Privacy guidelines should have steps for evaluation.
- ï Privacy requirements should be specified with engineering techniques such as anonymization, as developers find it difficult to relate privacy requirements into engineering techniques.

These recommendations would help addressing the issues we identified in this research. We intentionally disregarded concerns developers raised that related to their work environment when they practiced privacy as our study did not investigate the organizational constraints that affected developers privacy practices. Previous work that focuses on organizational privacy climate [14] has identified the importance of infrastructural support within the software development industry for developers to practice privacy when they develop software. Our study was designed to focus explicitly on the issues developers face as individuals when they are asked to embed privacy into software applications. When organizations provide infrastructural support and the environment for developers to practice privacy these issues needs to be addressed to make that support effective.

With the phenomenon of privacy engineering, translating vague privacy concepts into engineering solutions for software developers has become an attractive area of research [13]. The systematic approach for PIA proposed by Oetzel et al. [21] is one example which shows the development of privacy engineering solutions for developers to practice privacy within their development practices. However, privacy engineering as a phenomenon is still emerging and initiations similar to ours are important for the development of practical privacy engineering solutions that address the issues of software developers. Our work contributes to the knowledge that is needed by privacy engineers and the academia in order to develop effective privacy guidelines for software developers.

## 6 CONCLUSIONS AND FUTURE WORK

This research attempts to identify the issues faced by software developers when they attempt to embed privacy into software applications. Based on the findings of the study we derive guidelines to effectively support software developers when they attempt to embed privacy into software applications. However, the relatively small sample of participants in the study should be taken into account in generalizing our findings.

Our findings indicate that developers have practical issues when they attempt to embed privacy into software applications. Developers find it difficult to relate privacy requirements into engineering techniques and they lack knowledge on formally established privacy concepts such as PbD, which are well known in the domain of privacy research. Because of this the solutions researchers implement, expecting developers to be well versed with the privacy concepts may not work in software development environments. When developers lacked knowledge, their personal opinions and complex system requirements seem to take precedence over privacy requirements which eventually result in software applications with limited or no privacy embedded. Therefore we suggest that developers should be given formal education on privacy practices. More studies to formally evaluate developers' engagement and experience with existing privacy practices and identify personal opinions of developers that could affect their decisions in embedding privacy

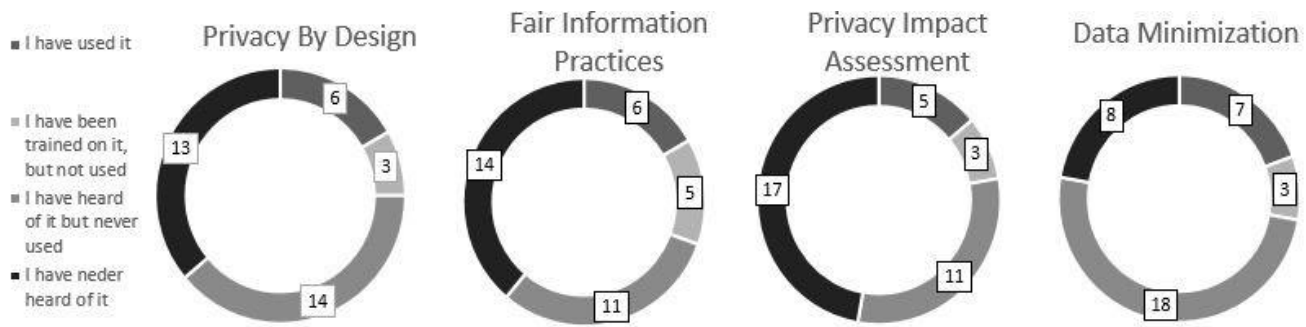

**Figure 1: Participants' Formal Knowledge on Privacy Concepts**

into software applications would be interesting avenues to continue this research.